\documentclass[10pt,a4paper,twocolumn,english,floatfix,groupedaddress,superscriptaddress,showpacs]{revtex4}

\usepackage{graphicx}
\usepackage{epsfig}
\usepackage[english]{babel}
\usepackage{amsmath}
\usepackage{amssymb}
\usepackage{amsfonts}
\usepackage{longtable}
\usepackage{dcolumn}
\usepackage{bm}
\usepackage{bbm}

\begin{document}
\title{Optimized Dynamical Decoupling for Power Law Noise Spectra}

\author{S. Pasini}
\email{pasini@fkt.physik.tu-dortmund.de}
\affiliation{Lehrstuhl f\"{u}r Theoretische Physik I, Dortmund University of 
Technology, Otto-Hahn Stra\ss{}e 4, 44221 Dortmund, Germany}
\author{G.S. Uhrig}
\email{goetz.uhrig@tu-dortmund.de}
\affiliation{Lehrstuhl f\"{u}r Theoretische Physik I, Dortmund University of 
Technology, Otto-Hahn Stra\ss{}e 4, 44221 Dortmund, Germany}

\date{\textrm{\today}}

\begin{abstract} 
We analyze the suppression of decoherence by means of dynamical decoupling 
in the pure-dephasing spin-boson model for baths with power law spectra. The 
sequence of ideal $\pi$ pulses is optimized according to the power of the 
bath. We expand the decoherence function and separate the cancelling 
divergences from the relevant terms. The proposed sequence is chosen to be the 
one minimizing the decoherence function. By construction, it provides 
the best performance. 
We analytically derive the conditions that must be satisfied. 
The resulting equations are solved
numerically. The solutions are  very close to the 
Carr-Purcell-Meiboom-Gill (CPMG) sequence for a soft cutoff of the bath while 
they approach the Uhrig dynamical-decoupling (UDD) sequence as the 
cutoff becomes harder. 
\end{abstract}


\maketitle

\section{Introduction}
\label{sec_Intro}

The dynamics of a spin $S=1/2$, or a quantum bit (qubit), coupled to an 
environment is one of the longest studied models of quantum decoherence. 
It finds important applications 
both in nuclear magnetic resonance (NMR) and in quantum information processing 
(QIP). The suppression of the decoherence is one of the goals one usually 
strives for. In order to achieve it, sequences of control pulses are used 
\cite{haebe76}.

There are two ways to address dynamical decoupling (DD) by control pulses.
Either the control is modulated continuously, see for instance Ref.\
 \onlinecite{gordo08}, or the control consists of short pulses 
which can be seen as (approximately) instantaneous. In this article, 
we will focus exclusively on the latter approach which is common
in NMR and wide-spread in QIP \cite{viola98,ban98,viola99a}. It  
relies on control pulses to invert the dynamics of the spin. The spin 
experiences a rotation of an angle $\pi$ at each pulse. The choice of the 
appropriate sequence is essential for an enhancement of the suppression 
of decoherence. A large variety of sequences has been suggested.
The majority is characterized by periodic pulses 
\cite{carr54,meibo58,viola98,ban98,viola99a,pryad06,zhang08}. 
The most famous example is the Carr-Purcell-Meiboom-Gill (CPMG) 
sequence \cite{carr54,meibo58} where cycles of two $\pi$ pulses are iterated.

 Other sequences proposed are
non-equidistant such as the concatenated dynamical decoupling (CDD) 
\cite{khodj05,khodj07,zhang08} or the Uhrig dynamical decoupling (UDD)  
\cite{uhrig07,uhrig08,uhrig09b}. The CDD consists of concatenations of 
pulse sequences. It can suppress   both transverse relaxation and 
longitudinal relaxation at the price of a relatively large number of 
pulses. If $\ell$ is the largest order in an expansion in 
the total duration  $t$,  in which no decoherence occurs, the required
number of pulses grows exponentially with $\ell$.
 The UDD eliminates only pure dephasing, but in turn it requires only a 
linearly growing number of pulses. The concatenation of the UDD sequence
(CUDD) allows for the suppression of transverse and longitudinal 
relaxation again at the price of an exponential growth of the number of pulses,
 but requiring only the square root of the number of pulses necessary
for CDD \cite{uhrig09b}.

We point out that all the above sequences are idealized 
in the sense that they are based on  
ideal, instantaneous pulses, i.e.,   $\delta$ peaks,
though the effect of finite pulse durations is being discussed. 
Furthermore, sequences of realistic pulses have been proposed 
\cite{skinn03,viola03,pryad08a,uhrig09c}. The finite duration of the 
pulse is a source of additional errors which can be reduced by designing 
 the shape of the pulses appropriately, see for instance
Ref.\ \onlinecite{pasin09a} and references therein.
In the present paper, however, we will concentrate only on sequences of 
instantaneous $\pi$ pulses.

The UDD sequence was discovered first for a spin-boson model 
\cite{uhrig07} where it was observed that no details of the model entered. 
On the basis of numerical evidence and finite order recursion
it was conjectured  that UDD is applicable
to any dephasing model \cite{lee08a,uhrig08}. This claim was
finally proven for arbitrary number of pulses
 in the total duration $t$ of the sequence \cite{yang08}. 
For various classical noise spectra the experimental verification of the 
theoretical results  was achieved   by optical 
control of the transition in Be ions \cite{bierc09a,bierc09b,uys09b}.
It was also shown 
that the UDD sequence  outperforms the CPMG sequence
and equidistant sequences in general for pure 
dephasing baths with hard cutoff while it performs worse for soft cutoffs
\cite{lee08a,uhrig08,bierc09a,bierc09b,uys09b}.  
This shows that the knowledge of the cutoff is an essential  piece of 
information for an optimum suppression of decoherence.

The performance of some pulse sequences for classical noise
spectra  has already been considered by Cywi\'{n}ski {\it et al.\ }
 for superconducting qubits subject to gaussian and random telegraph noise 
\cite{cywin08}. The authors compare the
 efficiency of different sequences in suppressing pure dephasing. They 
find that the UDD sequence is optimum in suppressing the decoherence
if the gaussian noise  displays a hard ultraviolet  (UV) cutoff.
In situations,  however, where one has to work in the regime of small 
frequencies (long times) such that the cutoff cannot be reached, 
the CPMG sequence is the one yielding the best  results. 

Another  proposal for an optimized sequence is put forward by Biercuk 
\textit{et al.\ } \cite{bierc09a,bierc09b}. 
The optimization in the UDD is extended 
to a locally optimized dynamical decoupling (LODD) sequence which is 
tailored to a given experimental noise environment. 
The experimental implementation of LODD for classical noise
shows that it performs better   than UDD and CPMG.
The LODD, however, is limited by the degree to which the spectral 
function of the noise (or the bath) is known.
 This caveat is dealt with by an optimized noise-filtration
dynamic decoupling (OFDD) \cite{uys09b}
where the noise spectrum is approximated by a constant up to 
a high-energy (UV) cutoff
for which sequences of pulses are deduced numerically.
For an ohmic bath the OFDD implies about the same factor of $\approx 1.5$
of improvement over the UDD sequence than the more cumbersome LODD.
But it does not provide a significant improvement for an ambient noise 
scaling such as $\propto 1/\omega^4$.

In this paper we generalize the OFDD from a constant spectrum to an
arbitrary power law without UV cutoff. This problem can be analysed
to a large extent analytically. The resulting \textbf{o}ptimized 
\textbf{d}ynamic \textbf{d}ecoupling 
sequences for \textbf{p}ower \textbf{l}aw spectra (PLODD) are universal
in the sense that the relative instants $\{\delta_j=t_j/t\}$ of the pulses 
depend only on the power law exponent $\alpha$.
The gist of our finding is that the PLODD sequences 
resemble CPMG sequences for slowly decreasing noise spectra while
they approach UDD sequences for fast decreasing noise spectra.
This agrees with the qualitative expectations based on other investigations 
\cite{cywin08,uhrig08,bierc09a,bierc09b,uys09b}.
Our results provide an important guideline in the choice of sequences to be 
applied in experiments on the suppression of decoherence.

From an experimental point of view, the noise used 
to test the dynamical decoupling sequences - classical noise, ohmic spectrum, 
as well as the $1/\omega^4$ noise spectrum - is in general governed by
 a power law spectrum. Analytically, power law spectra 
approximate any spectrum either for very small or for very large 
frequencies which in turn correspond to long or short durations, respectively.
Moreover, power law spectra are the perfect tool to analyze the 
influence of baths characterized by different cutoffs on the
optimized pulse sequence.
By varying the exponent of the power law of the spectrum we can simulate 
both baths with soft and hard cutoffs and we can interpolate smoothly
between them.

In order to render an analytical investigation possible we
consider the spin-boson model with pure dephasing. 
We start from the decoherence function $\chi(t)$ (defined in Eq.\ \eqref{chi})
which measures the size of the decoherence. The advantage is that the 
decoherence function  at the same time embodies both the characteristics of 
the bath and of the sequence of $\pi$ pulses. 
Hence one has to minimize $\chi(t)$.

The article is organized as follows. In Sect. \ref{sec_decoherence_function} 
the model is introduced and the decoherence function for a sequence of $\pi$ 
pulses is defined and  discussed. In the following Sect. 
\ref{sec_incomplete_gamma_function}, we study and solve the problem of the 
diverging terms in the integration yielding the decoherence function. 
Then we derive the final equations in Sect. \ref{sec_relevat_terms} 
and solve them numerically in Sect. \ref{sec_numerical_results}, where also 
examples of the PLODD are shown. At last the conclusions are drawn in Sect. 
\ref{sec_conclusion}.

\section{Convergence of the decoherence function}
\label{sec_decoherence_function}

We consider the spin-boson model with pure dephasing
\begin{equation}
 \label{spin_boson_model}
H=\sum_i\omega_i b^\dagger_i b_i+\frac{1}{2}\sigma_z\sum_i\lambda_i\left( 
b^\dagger_i+b_i\right)+E
\end{equation} 
describing a single qubit as a spin $S=1/2$ coupled linearly to a 
bosonic bath. The spin is represented by the Pauli matrix  $\sigma_z$, while 
the $b_i^{(\dagger)}$ are the annihilation (creation) operators of the bath. 
The constant $E$ sets the energy offset. The properties of the bath are 
defined by the set of parameters $\{\lambda_i,\omega_i\}$. This information is 
conveniently encoded in the spectral density \cite{legge87,weiss99} 
\begin{equation}
 \label{spectral_density} 
J(\omega) =\sum_i\lambda_i^2\delta(\omega-\omega_i).
\end{equation}

We recall that the quantum mechanical time evolution $p^n$
of a sequence with $n$ $\pi$ pulses about the $x$ axis of the
spin reads
\begin{equation}
\label{eq:udd1}
p^n = f_{t -\delta_{n}}X\, 
 f_{\delta_{n}-\delta_{n-1}}X\ldots
X\, f_{\delta_{3}-\delta_{2}} 
X\, f_{\delta_{2}-\delta_{1}} X \, 
f_{\delta_{1}},
\end{equation}
where $X$ stands for the spin operator of the rotation due
to the pulse \cite{yang08,uhrig09b}. Such a sequence
suppresses the relaxation along $z$  \cite{yang08,uhrig09b}. If $t$ is the 
total duration of the sequence, the instant $t_j$, at which the
pulse $j$ is applied, is given by $t_j=t\delta_j$.
By definition $\delta_0:=0$ and $\delta_{n+1}:=1$ although 
there is no pulse neither at the very beginning nor at the very end.
The notation $f_{\delta_{i}-\delta_{i-1}}$ stands for the free 
evolution of the system in the interval 
$t(\delta_{i}-\delta_{i-1})$ between two successive pulses.

In Refs.\ \onlinecite{uhrig07,uhrig08} it is shown that
the free induction decay is proportional to $e^{-2\chi(t)}$ where
the decoherence function is defined by
\begin{equation}
\label{chi}
\chi(t) := \int_0^\infty \frac{S(\omega)}{\omega^2}\lvert  
y_n(\omega t)\rvert ^2 \mathrm{d}\omega.
\end{equation} 
Here the noise spectrum $S(\omega)$ is related to the spectral density
$J(\omega)$ in \eqref{spectral_density}  by
\begin{equation}
 \label{S} 
S(\omega):=\frac{1}{4}J(\omega) \coth (\beta\omega/2),
\end{equation} 
where $\beta$ is the inverse temperature.
The filter function $y_n(z)$ ($z:=\omega t$) for $n$ pulses is given by
\begin{equation}
\label{y} 
y_n(z):=\sum_{j=0}^{n+1}2^{q_j} (-1)^je^{iz\delta_j},
\end{equation}
with
\begin{equation}
\label{q_j}
q_j:=\left\{\begin{array}{l}
0\ \mathrm{if}\ j=0,n+1\\
1\ \mathrm{if}\ j\in\{1,2,\dots ,n-1,n\}
\end{array}
\right. .
\end{equation}
Obviously, it encodes the properties of the sequence.

Equation (\ref{chi}) is the starting point for the evaluation of the optimized 
sequences. The aim is to keep $e^{-2\chi(t)}$ close to the unity 
as long as possible. In Ref.\ \onlinecite{uhrig07} the condition was enforced
that the first $n$ derivatives of the filter function should vanish at 
$\omega t=0$ for a sequence with $n$ pulses. Thus the function $y_n$ would 
increase very slowly close to zero. 
The condition on the derivatives 
implies  the following set of non-linear equations
\begin{equation}
\label{UDD_conditions}
0=\sum_{j=1}^{n+1}2^{q_j} (-1)^j\delta_j^p
\end{equation}
for $p\in\{1,2,...,n\}$. For $p=0$, Eq.\ (\ref{UDD_conditions}) is also zero 
\cite{uhrig07}, which is equivalent to $y_n(0)=0$. The solution 
of the Eqs.\ \eqref{UDD_conditions} reads \cite{uhrig07}
\begin{equation}
\label{UDD_delta} \delta_j^{\mathrm{UDD}}=\sin^2\left[j\pi/2(n+1)\right].
\end{equation} 
The UDD sequence suppresses the decoherence  best for baths with a hard 
cutoff rather than for baths with a very soft cutoff.
This was tested by means of numerical simulations \cite{lee08a}, analytical
analyses \cite{cywin08,uhrig08}, and 
experiments with classical noise \cite{bierc09a,bierc09a}.
 
\begin{figure}[ht]
    \begin{center}
    \includegraphics[width=0.99\columnwidth,clip]{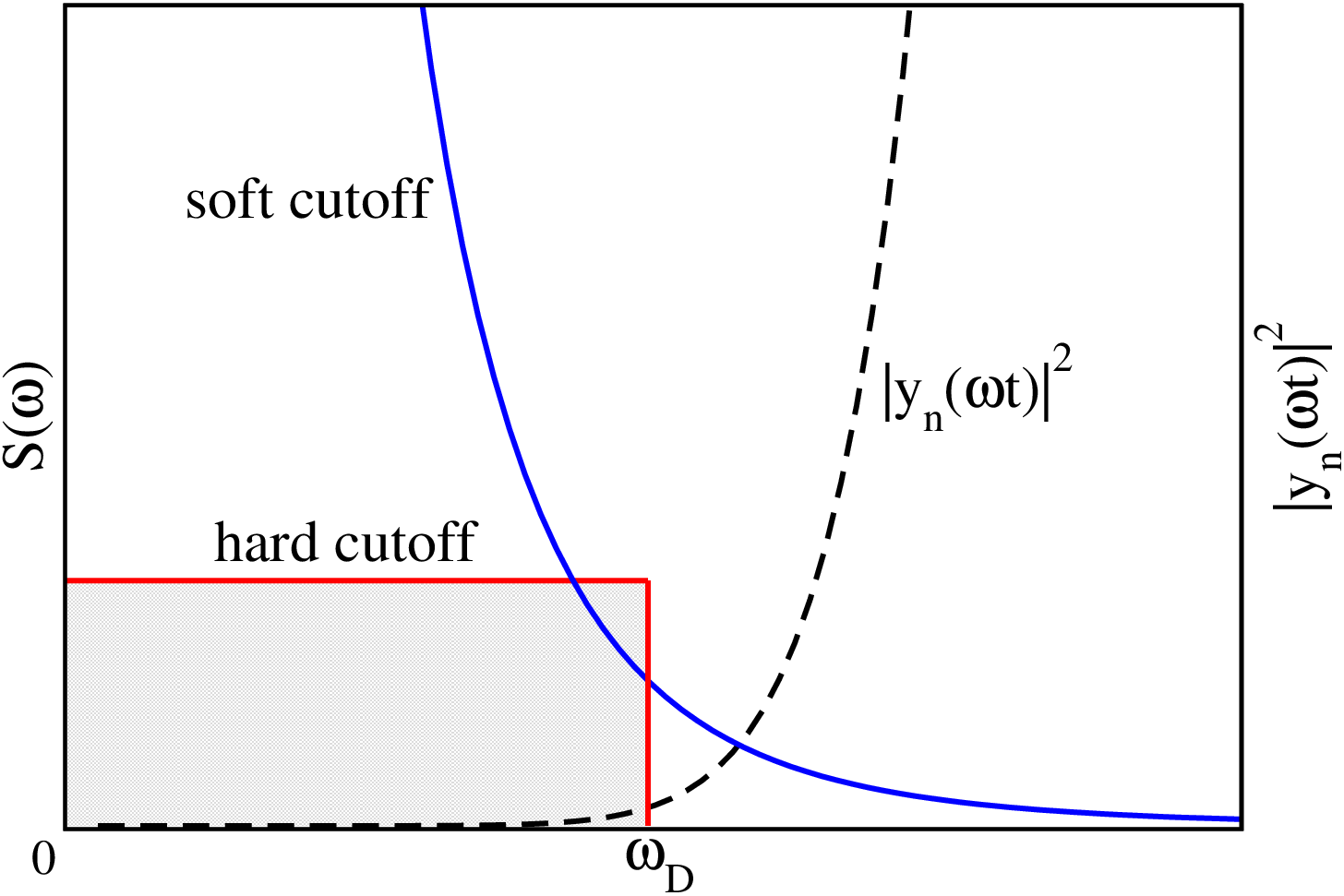}
    \end{center}
    \caption{(color online) 
     Schematic visualization of the filter function (black dashed line) and of 
     the spectral function $S(\omega)$ for a soft and for a hard cutoff. 
     $\omega_D$ stands for the ultraviolet cutoff. The plot illustrates that
     the overlap between the filter function and the spectral density  
     remains significant for a soft cutoff, even if $|y_n(\omega t)|^2$
     shifts to the right. This is in contrast to the situation for a hard 
     cutoff where the overlap becomes extremely small for 
     $n\to\infty$ or $t\to 0$.
      \label{fig0_spectral}
}
\end{figure}

The efficiency of UDD depends on the applicability of an expansion in powers of
 $t$ of $\chi(t)$ \cite{lee08a,uhrig08}. This expansion implies 
the  expansion of $y_n(z)$ in powers of $z=\omega t$. This is always possible
since $y_n(z)$ as defined in \eqref{y} is analytical. 
But the existence of the integrals of the resulting series
in powers of $\omega t$, as required by \eqref{chi}, depends
on the UV cutoff of $S(\omega)$ \cite{uhrig08}.

In the derivation of Eq.\ (\ref{UDD_delta}) only the existence of the 
derivatives of $y_n(\omega t)$ and 
not the existence of the integral over the frequency $\omega$ is required. The 
decoherence function consists of the product of the function $S(\omega)$ times 
the square modulus of the filter function. The function $\chi(t)$ is 
minimum if the overlap between $S(\omega)$ and $|y_n(z)|^2$ is minimum.
The significance of the UV cutoff is illustrated in Fig.\  \ref{fig0_spectral}.

From now on we focus on power law spectra
\begin{equation}
\label{powerlaw}
\frac{S(\omega)}{\omega^2} = \frac{S_0}{\omega^{\alpha+1}}.
\end{equation}
Then the decoherence function reads
\begin{equation}
\chi(t)=S_0\int_0^\infty \frac{1}{\omega^{\alpha+1}}
\lvert y_n(\omega t)\rvert^2 \mathrm{d}\omega
\end{equation}
It is required that $\alpha$ is  strictly positive to ensure the convergence 
for $\omega\to \infty$. 
This is true because the filter function is bounded from 
above $\lvert y_n(\omega t)\rvert\leqslant 2(n+1)$. 
The prefactor $S_0$ incorporates all the
 constants of the spectral density. 

For $\omega\to 0$ (infrared (IR) limit), $\chi(t)$ converges if 
$y_n(\omega t)\varpropto (\omega t)^m$ for $m$ large enough.
This in turn  depends on the choice of the sequence $\{\delta_j\}$. 
For arbitrary sequences $\{\delta_j\}$ we have
 $ y_n(0)=0$ so that the IR convergence is guaranteed for $\alpha<2$. 
For larger $\alpha$, we require  that 
the first $m$ derivatives of the filter function  vanish, implying
\begin{equation}
\label{y_vanish}
\lvert y_n(\omega t)\rvert ^2\varpropto(\omega t)^{2(m+1)},
\end{equation} 
which is similar, but not identical, to the requirement for the
UDD in Eq.\ \eqref{UDD_conditions}.  
The convergence of $\chi(t)$ is guaranteed for 
\begin{equation}
 \label{condition_alpha}
\alpha<2m+2.
\end{equation}

In Ref.\ \onlinecite{uhrig07} it was argued that UDD applies independently of 
the temperature. This indicates that UDD can be equally used to suppress 
classical gaussian noise \cite{cywin08,uhrig08}. 
For high temperature $\beta\to 0$ the thermal 
fluctuations dominate over the quantum fluctuations such that 
$S(\omega)\varpropto 1/\omega$ for $J(0)\neq 0$. This is the famous $1/f$ 
noise, which corresponds to $\alpha=2$ in our notation. 
The case $S(\omega) \propto 1/\omega^4$ is experimentally relevant
for ions in a Penning trap \cite{bierc09a,bierc09b}. This case
corresponds to $\alpha=5$ in the above notation.

The basic 2-pulse cycle of the CPMG sequence coincides with the UDD 
sequence for $n=2$. It makes  the first two derivatives of 
the filter function vanish \cite{uhrig07}. According to Eq.\ 
\ref{condition_alpha} its applicability is restricted to baths characterized 
by $\alpha < 6$.

In order to study general power laws we proceed as follows. 
We substitute  $z=\omega t$ in the decoherence function $\chi(t)$ in 
(\ref{chi}) obtaining
\begin{equation}
\label{chi_z}
 \chi(t)=S_0 t^\alpha I_n
\end{equation} 
with
\begin{equation}
\label{In}
 I_n:=\int_0^\infty \frac{\lvert y_n(z)\rvert ^2}{z^{\alpha+1}}\mathrm{d}z.
\end{equation}
This simple substitution reveals that the optimum $\{\delta_j\}$ are 
independent of the total duration $t$ of the sequence. 
All the time-dependence of $\chi(t)$ is a simple power of $t$ as
in \eqref{chi_z}. Its exponent is determined by the power law spectrum 
of the bath. The precise sequence $\{\delta_i\}$ determines the factor $I_n$.

The condition for the first $m$ derivatives to vanish is  given by the 
set of the first 
$m$ non-linear equations in \eqref{UDD_conditions}, i.e.,
for $p\in\{1,2,\ldots, m\}$. These conditions 
are the same as those leading to the UDD sequence \cite{uhrig07}
except that they do not need to be fulfilled up to $p=n$
but only up to $p=m$. For a sequence of $n>m$ pulses we still have $n-m$ 
degrees of freedom left. This freedom is used  to  minimize $I_n$ and hence 
the decoherence function $\chi(t)$.
In this minimization the $m$  Eqs.\
(\ref{UDD_conditions}) act as additional constraints. 
Hence,  we have to study the variation
\begin{equation}
\label{PLODD_condition}
\frac{\partial}{\partial \delta_j}
\left[I_n-\sum_{i=1}^m\lambda_i\left. 
\frac{\partial y_n(z)}{\partial z}\right|_{z=0}\right]=0,
\end{equation} 
where $m$ Lagrange multipliers $\lambda_i$ appear due to the $m$ constraints.

\section{Diverging terms}
\label{sec_incomplete_gamma_function}

The integral $I_n$ (\ref{In}) converges if the condition $\alpha<2(m+1)$
is fulfilled. But the integration in \eqref{chi} cannot be carried
out analytically. To make analytical progress we split the square modulus
of the filter function into a sum of exponential terms according to
\begin{equation}
 \label{y_modulus} 
|y_n(z)|^2=
\sum_{i,j=0}^{n+1}2^{q_i+q_j}(-1)^{i+j}e^{iz\Delta_{ij}},
\end{equation} 
where we use the
 notation $\Delta_{ij}:=i(\delta_i-\delta_j)-0^+$ for $\{i,j\}=0..n+1$;
the summand $-0^+$ stands for an infinitesimal negative real part
which is required later on for convergence for $z\to+\infty$.
The integral $I_n$ is given by the limit of the sum
\begin{subequations}
\begin{eqnarray}
 I_n &=&\lim_{x\to0+}I_n(x)
\\
\label{In_vs_Iij}
I_n(x) &:=&
\sum_{i,j=0}^{n+1}2^{q_i+q_j}(-1)^{i+j}I_{ij}(x)
\end{eqnarray}
\end{subequations}
where the integrals
\begin{equation}
\label{Iij}
 I_{ij}(x):=\int_x^\infty \frac{e^{\Delta_{ij}z}}{z^{\alpha+1}}\mathrm{d}z.
\end{equation} 
 UV convergence is ensured by the infinitesimal negative
real part of $\Delta_{ij}$, see definition below Eq.\ \eqref{y_modulus}.

The regularization by a finite IR cutoff $x$ is required because
the limit $x\to 0$ does not exist for the individual terms $I_{ij}(x)$.
Each term $I_{ij}(x)$ can be reduced to an analytical expression
by the substitution  $z\to -z/\Delta_{ij}$
\begin{subequations}
\begin{eqnarray}
 \label{Iij_x}
I_{ij}(x)&=&(-\Delta_{ij})^\alpha\int_{-x\Delta_{ij}}^\infty 
\frac{e^{-z}}{z^{\alpha+1}}\mathrm{d}z
\\
&=& (-\Delta_{ij})^\alpha \Gamma(-\alpha,-\Delta_{ij}x)
\label{incomplete_gamma}
\end{eqnarray} 
\end{subequations}
where $\Gamma(-\alpha,-\Delta_{ij}x)$ is the incomplete Gamma 
function \cite{abram64}.

For later use we state that the vanishing of the first
$m$ derivatives of the filter functions $y_n(z)$ at $z=0$  implies
\eqref{y_vanish} and thus
\begin{subequations}
\begin{eqnarray}
\left.\partial_z^{p}|y_n(z)|^2\right|_{z=0}&=&
\sum_{ij=0}^{n+1}2^{q_i+q_j}(-1)^{i+j}\left(\Delta_{ij}\right)^p
\\
&\equiv& 0
 \label{dy_vs_Delta_null} 
\end{eqnarray} 
\end{subequations}
for $0\leqslant p \leqslant 2m+1$. Eq.\ (\ref{dy_vs_Delta_null}) 
asserts that the weighted sum of powers of $\Delta_{ij}$ vanishes.
We will utilize this cancellation to find the relevant 
contributions to $I_n$ analytically.

\subsection{Expansion of the incomplete Gamma function}
\label{subsec_expansion_gamma}

\paragraph{Non-Integer Exponents}
Starting from the definition of the incomplete Gamma function $\Gamma(a,x)$ 
with $-a = \alpha \notin {\mathbb N}_0$ we integrate by parts 
$\ell$ times and write
\begin{equation}
\label{Gamma_T1_T2}
 \Gamma(a,x)= {\cal T}_1+ {\cal T}_2.
\end{equation} 
with
\begin{eqnarray}
\label{gamma_T1}
 {\cal T}_1=-\mathrm{e}^{-x}\sum_{p=0}^{\ell-1}x^{a+p}\frac{(a-1)!}{(a+p)!}
\end{eqnarray}
and
\begin{eqnarray}
 \label{gamma_T2}
{\cal T}_2=\frac{(a-1)!\ \Gamma(a+\ell,x)}{(a+\ell-1)!}.
\end{eqnarray}

For $a\leqslant 0$ the limit $\lim_{x\to 0}\Gamma(a,x)$ is not defined. We 
choose $\ell$ such that 
\begin{equation}
\label{ell_choice}
a+\ell > 0 > a+\ell-1.
\end{equation}
For non-integer $a$ this is possible. The first inequality ensures that 
$\lim_{x\to 0}\Gamma(a+\ell,x)=\Gamma(a+\ell)$. 
The recurrence relation $\Gamma(z+1)=z\Gamma(z)$ implies
\begin{equation}
 \lim_{x\to 0}{\cal T}_2=\Gamma(a).
\end{equation}
Now we concentrate on the term ${\cal T}_1$. Since the exponential function 
$\mathrm e^{-x}$ can be expanded in powers of $x$, ${\cal T}_1$ has a 
well-defined expansion
\begin{equation}
{\cal T}_1(x)=x^a\sum_{p=0}^\infty \alpha_p x^p.
\end{equation} 
The coefficients $\{\alpha_p\}$ depend on the coefficients of ${\cal T}_1$
and on those of the expansion of the exponential. Their explicit 
form does not matter here. The powers diverge
 in the limit  $x\to 0$ for $p\in\{0,\dots,\ell-1\}$.
For $p\geqslant \ell$  they vanish for $x\to 0$. 

Finally we write the integral $I_n(x)$ in terms of the integrals $I_{ij}(x)$ 
defined in Eq.\ (\ref{Iij}) and evaluated in  Eq.\ (\ref{incomplete_gamma})
\begin{subequations}
\label{noninteger}
\begin{eqnarray}
\label{In_noninteger_1}
 I_n(x)&=&\sum_{p=0}^{\ell-1}\alpha_p x^{p-\alpha}
\sum_{i,j=0}^{n+1}2^{q_i+q_j}(-1)^{i+j}(-\Delta_{ij})^p
\qquad \\ 
\label{In_noninteger_2}
&+&\sum_{i,j=0}^{n+1}2^{q_i+q_j}(-1)^{i+j}(-\Delta_{ij})^\alpha
\Gamma(-\alpha)
 \\
&+& {\cal O}(x^{\ell-\alpha}).
\label{In_noninteger_3}
\end{eqnarray} 
\end{subequations}
The last  term ${\cal O}(x^{\ell-\alpha})$ 
comprises all the contributions which vanish for $x\to 0$, recall
\eqref{ell_choice} for $a=-\alpha$.
The first term \eqref{In_noninteger_1} contains the diverging terms. 
But they cancel one another completely because the inner sum
in \eqref{In_noninteger_1} vanishes due to \eqref{dy_vs_Delta_null}.
To see this one must use \eqref{condition_alpha} and \eqref{ell_choice} 
to arrive at $p\le \ell -1 \le 2m+1$.
This rather formal argument simply reflects the fact that
\eqref{condition_alpha} guarantees the IR convergence of the
integration in \eqref{chi}. Hence all IR divergent terms appearing in 
intermediate calculations  have to cancel finally.

From the above the only remaining and thus relevant
contribution to $I_n=\lim_{x\to0+}I_n(x)$ is \eqref{In_noninteger_2} 
\begin{equation}
 \label{In_noninteger_final} 
I_n=\Gamma(-\alpha)
\sum_{i,j=0}^{n+1}2^{q_i+q_j}(-1)^{i+j}(-\Delta_{ij})^\alpha.
\end{equation}
This is the result for non-integer $\alpha$.
Note that $\Gamma(-\alpha)$ is only a global 
prefactor which does not depend on $\Delta_{ij}$.

\paragraph{Integer Exponents}
Next we consider the case of $\alpha={\mathbb N}_0$. 
We use Eqs.\ (\ref{gamma_T1},\ref{gamma_T2}) with $\ell=\alpha=-a$
\begin{eqnarray}
 \label{gamma_n_eq_alpha} 
\Gamma(-\alpha,x)=&-&{\mathrm e}^{-x}
\sum_{p=0}^{\alpha-1}x^{p-\alpha}(-1)^{1+p}\frac{(\alpha-1-p)!}{\alpha!}
\nonumber \\
&+& \frac{(-1)^\alpha}{\alpha!}\Gamma(0,x).
\end{eqnarray} 
Thus Eq.\ \eqref{noninteger} now reads
\begin{subequations}
\label{integer}
\begin{eqnarray}
\label{In_integer_1}
&& I_n(x) = \sum_{p=0}^{\alpha-1}\alpha_p x^{p-\alpha}
\sum_{i,j=0}^{n+1}2^{q_i+q_j}(-1)^{i+j}(-\Delta_{ij})^p
 \\ 
\label{In_integer_2}
&&+\frac{(-1)^\alpha}{\alpha!}
\sum_{i,j=0}^{n+1}2^{q_i+q_j}(-1)^{i+j}(-\Delta_{ij})^\alpha
\Gamma(0,-\Delta_{ij} x).\quad\quad\
\end{eqnarray} 
\end{subequations}
As before the inner sum in \eqref{In_integer_1} cancels because
of \eqref{dy_vs_Delta_null} and because \eqref{condition_alpha}
implies $p\leq \alpha -1 \leq 2m$. Note that \eqref{condition_alpha}
and \eqref{dy_vs_Delta_null}
additionally imply that the weighted sum of the powers 
$\Delta_{ij}^\alpha$ vanishes. This will help
to simplify \eqref{In_integer_2} further in the limit $x\to 0$.
We  expand
the incomplete Gamma function
\begin{equation}
\label{gamma_expansion} 
\Gamma(0,-\Delta_{ij}x)=-\gamma-\ln (x)-\ln(-\Delta_{ij})+{\cal O}(x),
\end{equation}
where $\gamma$ is the Euler-Mascheroni constant.
Hence in the limit  $x\to 0$ the only non-vanishing 
contribution to $I_n=\lim_{x\to 0}I_n(x)$ is
\begin{equation}
\label{prefinal_Integer}
I_n = \frac{-1}{\alpha!}
\sum_{i,j=0}^{n+1}2^{q_i+q_j}(-1)^{i+j}\Delta_{ij}^\alpha
\ln(-\Delta_{ij}).
\end{equation}

\subsection{Example for $\alpha=2$}

For $\alpha=2$ we explicitly write the expansion of the incomplete Gamma 
function and show that the diverging terms cancel.
For $x\to 0$ we have
\begin{subequations}
\begin{eqnarray}
 \label{Gamma_alpha2}
\Delta_{ij}^2 \Gamma(-2,\Delta_{ij}x)&=&
\frac{\Delta_{ij}^2}{4}(3-2\gamma-2\ln(-\Delta_{ij}x))\nonumber
\\
&&+\frac{1}{2x^2}+\frac{\Delta_{ij}}{x}+{\cal O}(x) 
\\
&=&I_{ij}^{(-2)}(x)+I_{ij}^{(-1)}(x)+I_{ij}^{(0)}\nonumber 
\\
&&-\frac{\Delta_{ij}^2}{2}\ln(-\Delta_{ij})+{\cal O}(x).
\end{eqnarray}
\end{subequations}
The function $I_n^{(k)}(x)$ is the sum of the $I_{ij}^{(k)}(x)$ 
weighted according to the right hand side of (\ref{In_vs_Iij}).
These contributions vanish. For $I_n^{(-2)}$ we can  write
\begin{subequations}
\begin{eqnarray}
I_n^{(-2)}(x)&=&\frac{1}{2x^2} \sum_{i,j=0}^{n+1}2^{q_i+q_{j}}(-1)^{i+j}
\\
\label{middle}
&=&\frac{1}{2x^2} \sum_{i=0}^{n+1}2^{q_i}(-1)^{i}
\sum_{j=0}^{n+1}2^{q_{j}}(-1)^{j}\\
&=&0.
\end{eqnarray} 
\end{subequations}
The vanishing of \eqref{middle} is guaranteed by Eq.\ 
(\ref{UDD_conditions}) for $p=0$ or, equivalently, by the property $y_n(0)=0$.
Because of the sum over all $\{ij\}$ we have
\begin{subequations}
\begin{eqnarray}
I_n^{(-1)}(x)&=&\frac{i}{x} \sum_{ij=0}^{n+1}2^{q_i+q_j}(-1)^{i+j}
(\delta_i-\delta_j)
\\
&=&\frac{i}{x} \sum_{j=0}^{n+1} 2^{q_{j}}(-1)^{j}
\sum_{i=0}^{n+1}2^{q_i}(-1)^{i}\delta_i
\nonumber
\\
&-&\frac{i}{x}\sum_{i=0}^{n+1} 2^{q_{i}}(-1)^{i}
\sum_{{j}=0}^{n+1}2^{q_{j}}(-1)^{j}\delta_{j}\\
&=&0.
\end{eqnarray} 
\end{subequations}
Similarly one obtains for $I_n^{(0)}$
\begin{eqnarray}
\nonumber
 I_n^{(0)} (x)&=&\frac{2\gamma+2\ln(x)-3}{4}
\sum_{ij=0}^{n+1}2^{q_i+q_j}(-1)^{i+j}(\delta_i-\delta_j)^2
\\
&=&0.
\end{eqnarray}
The last equation vanishes because of Eq.\ (\ref{UDD_conditions}) at 
$p=0$ or at $p=1$. This is because 
$(\delta_i-\delta_j)^2=\delta_i^2+\delta_j^2-2\delta_i\delta_j$.
 Eq.\ (\ref{UDD_conditions}) has to hold for $p=0$ and $p=1$
because these values fulfill $p\leq m$  since
$m\geq 1$ is required by  (\ref{condition_alpha}) for $\alpha=2$.

\section{Relevant terms and final equations}
\label{sec_relevat_terms}

Eqs.\ (\ref{In_noninteger_final},\ref{prefinal_Integer}) 
provide the analytical results for $I_n$. One further
simplification  stems from the fact that $I_n$ is real.
Hence only the real parts of the summands in 
(\ref{In_noninteger_final},\ref{prefinal_Integer})
need to be included since the imaginary parts cancel. 
 We analyze the case $\alpha\in {\mathbb N}_0$, where  
we distinguish even and odd exponents, and the 
case $\alpha\notin {\mathbb N}_0$ separately.

\subsection{Integer Exponent}
\paragraph{$\alpha$ even.} 
We use $\Delta_{ij}:=i\varphi_{ij}-0^+$ with 
$\varphi_{ij}=\delta_i-\delta_j$ and
 $\ln(r{\mathrm e}^{i\theta})=\ln(r)+i\theta$ with $r>0$
and $|\theta|<\pi$  to obtain
\begin{eqnarray}
 \label{even_T2} 
\mathrm {Re}\, \Delta_{ij}^\alpha\ln(-\Delta_{ij}) 
&=&(-1)^{\alpha/2}
\varphi_{ij}^\alpha \mathrm {Re}\left(\ln|\varphi_{ij}|-i
\frac{\pi}{2}\mathrm{sgn}\varphi_{ij}\right)
\nonumber \\
&=& (-1)^{\alpha/2}\varphi_{ij}^\alpha\ln|\varphi_{ij}|.
\end{eqnarray} 
 The sum over all $i\neq j$ as required by \eqref{prefinal_Integer} reads
\begin{eqnarray}
\label{In_alpha_even}
 && I_n^{\mathrm{even}} = \frac{(-1)^{1+\frac{\alpha}{2}}}{\alpha!}
\nonumber \\
&&\times
\sum_{\begin{subarray}{c} i,j=0 \\ 
i\neq j\end{subarray}
}^{n+1}2^{q_i+q_j}(-1)^{i+j}|\delta_i-\delta_j|^\alpha\ln|\delta_i-\delta_j|.
\qquad
\end{eqnarray}

\paragraph{$\alpha$ odd.} Starting from
\begin{eqnarray}
 \label{odd_T2} 
&&\mathrm {Re}\, \Delta_{ij}^\alpha\ln(-\Delta_{ij})  
 = \nonumber  \\
&& \qquad (-1)^{\frac{\alpha-1}{2}} \varphi_{ij}^\alpha \mathrm {Re}
\left[ i\left(\ln|\varphi_{ij}|
-i\frac{\pi}{2}\mathrm{sgn}\varphi_{ij}\right)\right]
\nonumber\\
&& \qquad = (-1)^{\frac{\alpha-1}{2}}
|\varphi_{ij}|^\alpha\frac{\pi}{2}.
\end{eqnarray}
the final integral $I_n$ becomes
\begin{eqnarray}
\label{In_alpha_odd}
I_n^\mathrm{odd}&=&\frac{(-1)^{\frac{\alpha+1}{2}}}{\alpha !}
\frac{\pi}{2} 
\nonumber \\
&&\times
\sum_{\begin{subarray}{c} i,j=0 \\ 
i\neq j\end{subarray}
}^{n+1}2^{q_i+q_j}(-1)^{i+j}|\delta_i-\delta_j|^\alpha.
\qquad
\end{eqnarray}

\subsection{$\alpha$ Non-Integer}
For positive non-integer $\alpha$ we consider 
\begin{subequations}
\begin{eqnarray}
 \label{notint_T2} 
\mathrm {Re}\, (-\Delta_{ij})^\alpha &=& |\varphi_{ij}|^\alpha\mathrm{Re} 
\left({\mathrm e}^{-i(\pi/2)\alpha\ \mathrm{sgn}\varphi_{ij}}\right)
\\
&=& \cos((\pi/2)\alpha) |\varphi_{ij}|^\alpha,
\end{eqnarray} 
 \end{subequations}
which implies
\begin{eqnarray}
\label{In_alpha_notint}
I_n^\mathrm{ni}&=& \cos((\pi/2)\alpha)\Gamma(-\alpha)
\nonumber \\
&&\times
\sum_{\begin{subarray}{c} i,j=0 \\ 
i\neq j\end{subarray}
}^{n+1}2^{q_i+q_j}(-1)^{i+j}|\delta_i-\delta_j|^\alpha.
\qquad
\end{eqnarray} 
where the superscript \textquoteleft ni\textquoteright \ 
of $I_n^\mathrm{ni}$ stands for \textquoteleft non-integer\textquoteright.

\begin{figure}[ht]
    \begin{center}
    \includegraphics[width=0.99\columnwidth,clip]{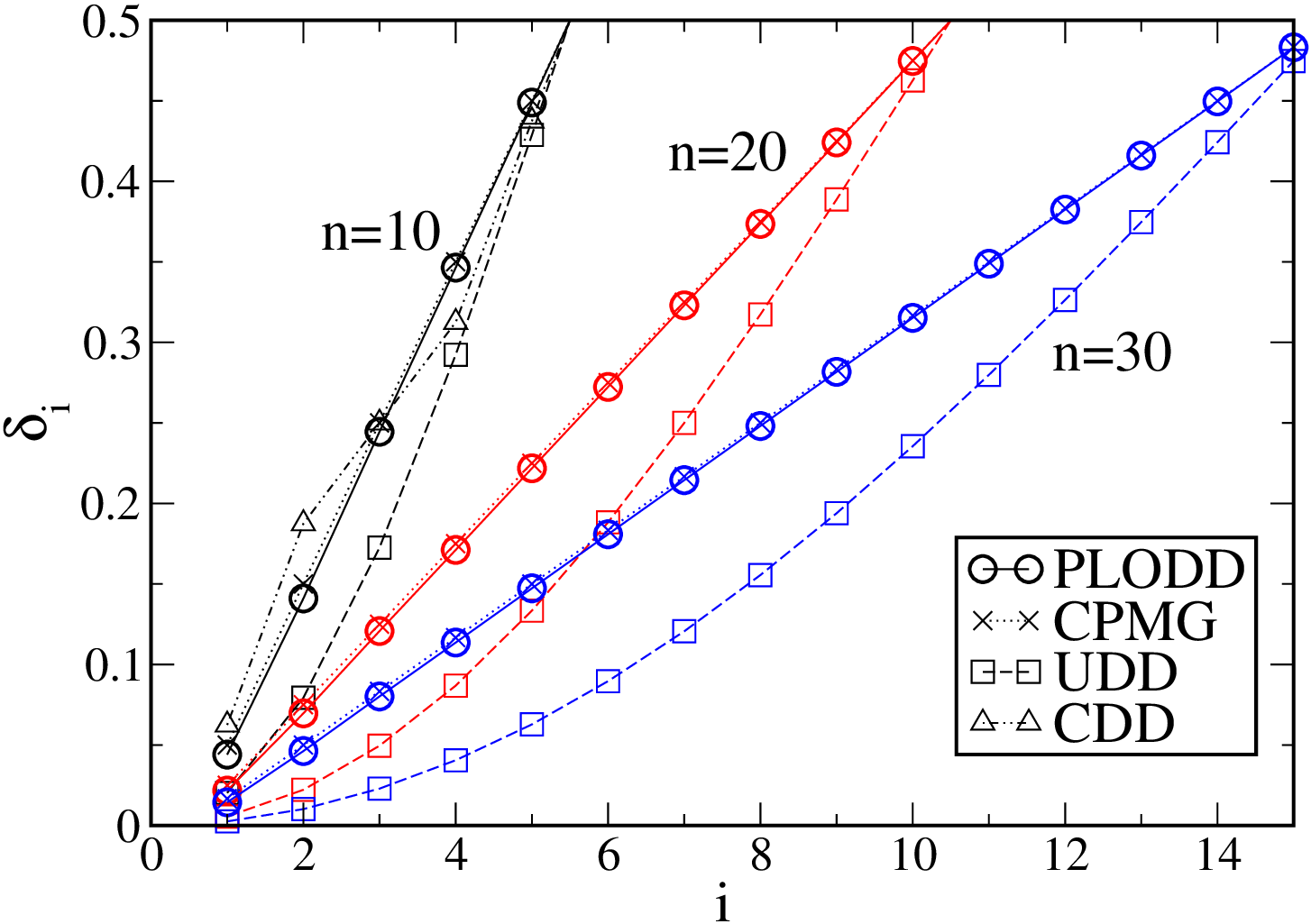}
    \end{center}
    \caption{(Color online) 
     Comparison of the optimized sequences PLODD and UDD and the sequence 
     CPMG for $\alpha=4$ and for various numbers of pulses $n$.
     For clarity only the first half of the sequences is plotted but the 
     symmetry about $\delta=0.5$ ($\delta_{n+1-j}=1-\delta_j$) 
     allows for the straightforward 
     reconstruction of the second half. The concatenated 
     sequence CDD is shown only for $n=10$ because it does not exist
     for $n=20$ or $n=30$. Note that we refer here  to the 
     CDD for pure dephasing \cite{khodj07}.
      \label{fig1_alpha4}
}
\end{figure}
\begin{figure}[ht]
    \begin{center}
    \includegraphics[width=0.99\columnwidth,clip]{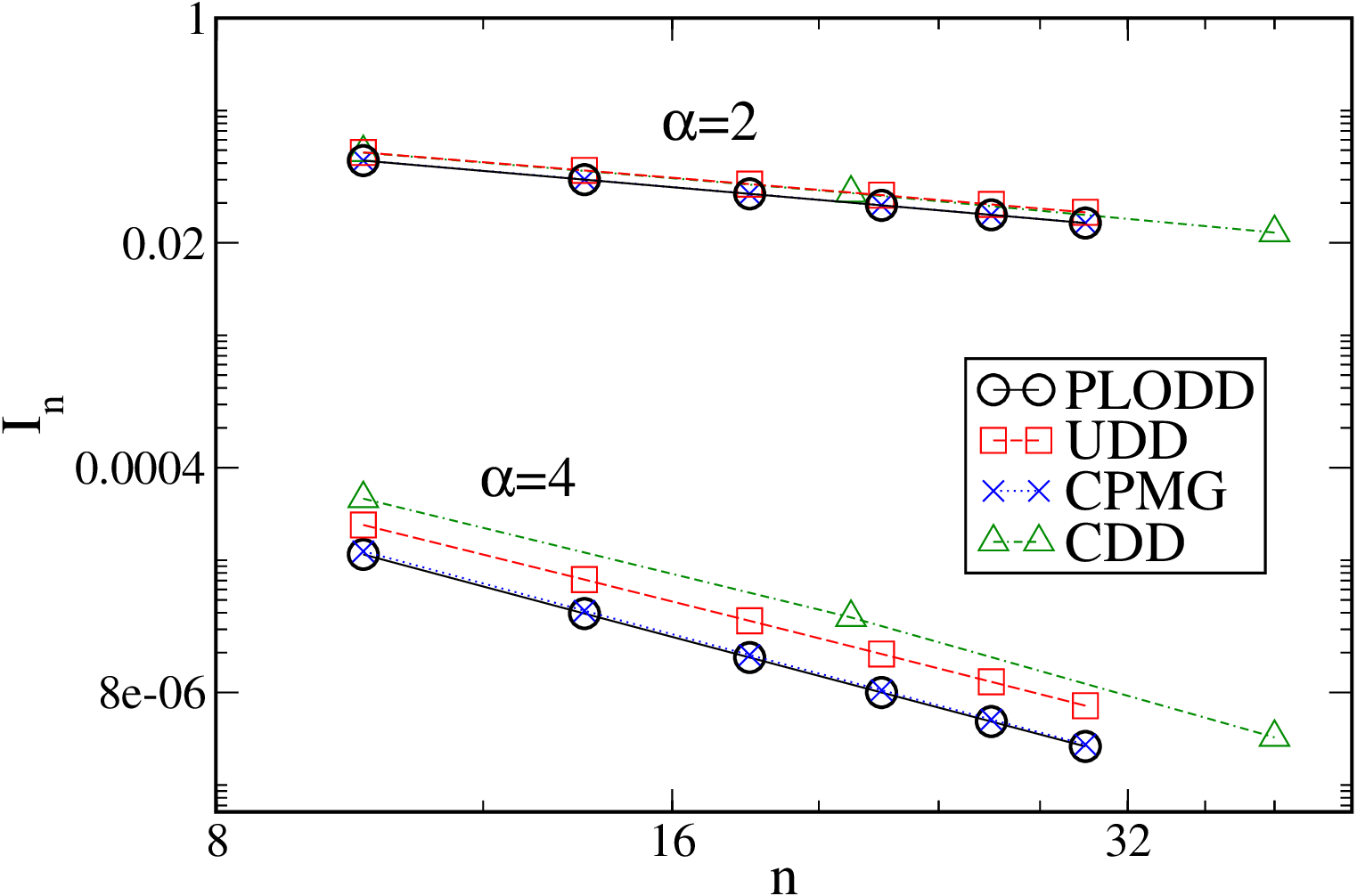}
    \end{center}
    \caption{(Color online)
      Log-log plot of the integral $I_n$ as a function of the total number of 
      pulses $n$ for $\alpha=2$ and $\alpha=4$. 
      Different sequences are compared, including the CDD sequence
      for pure dephasing  \cite{khodj07}. 
      The latter takes only the values $n=\{0, 1, 2, 5, 10, 21, 42, \ldots \}$.
      \label{fig_In_vs_n}
}
\end{figure}

\section{Numerical results}
\label{sec_numerical_results}

The power law optimized dynamical decoupling (PLODD) is a bath-optimized 
sequence depending on the exponent $\alpha$ only. For various values of 
$\alpha$, we  numerically solve the system of non-linear equations 
\begin{subequations}
  \label{PLODD_eqs}
  \begin{eqnarray} 
    \label{plodd-udd}
    0&=& 
    \sum_{j=1}^{n+1}2^{q_j} (-1)^j\delta_j^p 
    \\
    \label{plodd-opt}
    0 &=& 
    \frac{\partial}{\partial \delta_j}\left[I_n-\sum_{i=1}^{[\alpha/2]}
      \lambda_i\left. 
      \frac{\partial y_n(z)}{\partial z}\right|_{z=0}\right]
  \end{eqnarray}
\end {subequations}
where $p\in\{1,...,\left[\frac{\alpha}{2}\right]\}$; here
$[x]$ stands for the largest integer not larger than $x\in\mathbb{R}$
and $[\alpha/2]$ results from (\ref{condition_alpha}).
The prefactor $I_n$ is computed analytically
 both in the case of integer and of 
non-integer values of $\alpha$; it is given in Eqs.\
(\ref{In_alpha_even},\ref{In_alpha_odd},\ref{In_alpha_notint}). 

In the sequel, we restrict ourselves to  symmetric sequences for two reasons.
First, the main results in dynamical decoupling are derived for 
symmetric sequences. Second, we  searched for asymmetric optimized
sequences for small number of pulses, but those sequences
found did not  perform better than the symmetric ones.
Hence we focus on symmetric sequences fulfilling $\delta_{n+1-j}=1-\delta_j$. 
In this case one has to deal with a system of $n/2+[\alpha/2]$ equations, 
solved for $n/2$ variables $\{\delta_i\}$ and $[\alpha/2]$ Lagrange multipliers
$\lambda_m$. For example, for $\alpha=4$ the system consists of 
$n/2+2$ equations. 

Figure \ref{fig1_alpha4} shows the resulting PLODD sequences
 for $\alpha=4$ for various number of pulses $n$. The PLODD sequences 
are very close to the CPMG ones. We recall that 
$\delta_j^{\mathrm{CPMG}}=(2j-1)/2n$. No relevant dependence of the PLODD 
instants as functions of the number of pulses $n$ can be observed. The 
concatenated sequence CDD for pure dephasing 
is also shown for comparison for $n=10$. We 
recall its recursion 
$p^{\mathrm{CDD}}_{n+1}=p_{n}^{\mathrm{CDD}} X p_{n}^{\mathrm{CDD}}$
for $n$ even while 
$p^{\mathrm{CDD}}_{n+1}=p_{n}^{\mathrm{CDD}} p_{n}^{\mathrm{CDD}}$
holds for $n$ odd; $p_0$ stands for the free evolution without pulse.

In Fig.\ \ref{fig_In_vs_n} the evaluation of the prefactor $I_n$ shows that 
PLODD performs slightly better than CPMG for $\alpha=4$ while for $\alpha=2$
 the data for PLODD and for CPMG coincide. It is interesting to notice how the 
performance changes with $\alpha$. For $\alpha=2$ we see that CDD and UDD 
provide almost the same results while for $\alpha=4$ UDD performs better than
CDD, though still outperformed by  CPMG and PLODD.
 In addition, Fig.\ \ref{fig_In_vs_n} 
indicates that $I_n$ decreases if $\alpha$ increases. We will 
come back to this point below.

The log-log plot in Fig.\ \ref{fig_In_vs_n} shows that $I_n$ scales 
like a power law in $n$. The regression $\ln I_n=a_1\ln n+a_0$  yields 
$a_0=-(2.33\pm 0.01)$ and $a_1=-(3.041\pm 0.003)$ for PLODD and 
$a_0=-(2.231\pm 0.009)$ and $a_1=-(3.062\pm 0.003)$ for CPMG. We can compare 
these results with Eq.\ (25) in Ref.\ \onlinecite{cywin08}
derived by Cywi\'{n}ski \textit{et al.}. 
Cywi\'{n}ski's  formula was derived for a {\it de facto}
infinite UV cutoff for $1.5\leqslant\alpha\leqslant2.5$ 
(or $0.5\leqslant\alpha_{\mathrm{cyw}}\leqslant 1.5$ in the notation
of Ref.\  \onlinecite{cywin08}). It shows that $\chi(t)$ 
scales like $n^{\alpha-1}$.
 
As a further check we calculated $I_n$  versus $n$ for CPMG 
for $1/f$ noise, i.e., $\alpha=2$. We find $a_1=-(0.9899\pm 0.0008)$ and 
$a_0=-(0.199\pm 0.002)$, which is close to the corresponding value 
$\ln C_1\simeq -0.163$  reported in Ref.\ \onlinecite{cywin08}. 
Cywi\'{n}ski \textit{et al.\ } concluded that in the range 
$0.5\leqslant\alpha_{\mathrm{cyw}}\leqslant 1.5$ CPMG is to be preferred over
UDD for the prolongation of  qubit coherence. 
The UDD outperforms the other sequences in the range where a finite UV cutoff 
makes itself felt.
Our systematic minimization confirms the results 
by Cywi\'{n}ski \textit{et al.}. It extends them
by putting them on a systematic basis leading to 
 the optimum power law dynamic decouling and
 because  a larger range of exponents is treated.

\begin{figure}[ht]
    \begin{center}
    \includegraphics[width=0.99\columnwidth,clip]{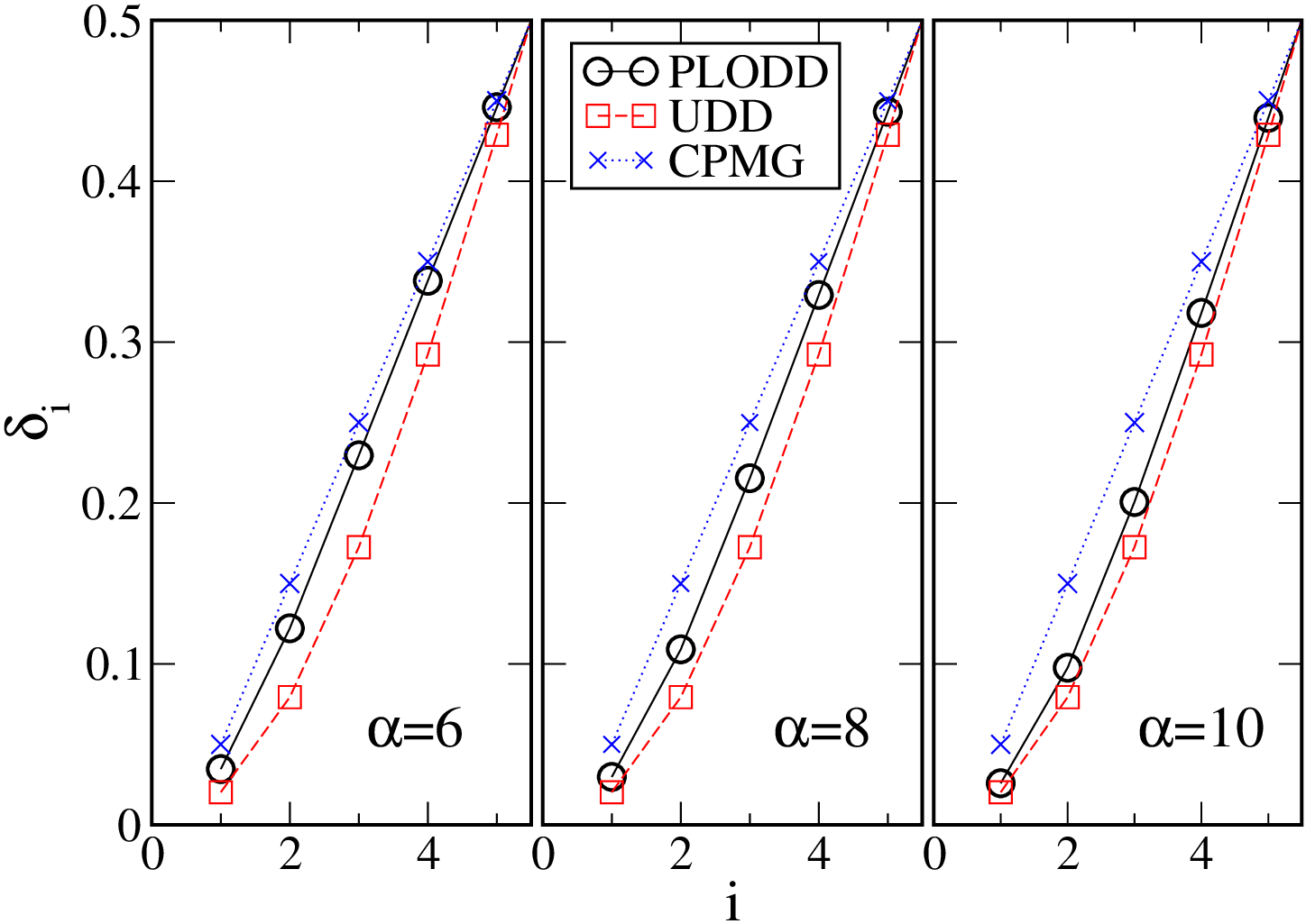}
    \end{center}
    \caption{(Color online) 
     Evaluation of PLODD sequences  for various exponents for $n=10$. 
     For clarity only the $\delta_i$ from $i=1$ to 5 are
     plotted. The second half of the sequence is symmetric to the first one
     ($\delta_{n+1-j} =1 -\delta_j$). 
     The PLODD approaches UDD for increasing $\alpha$. 
 \label{fig2_n10}
}
\end{figure}

If we consider higher values of $\alpha$,  the PLODD 
sequence approaches the UDD one. This is illustrated in Fig.\ \ref{fig2_n10}. 
The instants of the PLODD sequence appear to be bounded from below by the 
instants of the CPMG sequence and from above by the those of the UDD one. For 
$\alpha\geq 6$ 
CPMG does not satisfy the condition (\ref{condition_alpha}) anymore
which is required for the 
convergence of $\chi(t)$ so that no comparison to PLODD or UDD is possible. 
For small values of $\alpha$, $[\alpha/2]$ is also small. 
This implies that the conditions \eqref{plodd-udd} are less important 
than the conditions  \eqref{plodd-opt} resulting from the  minimization of the 
prefactor $I_n$. This finding confirms what was already 
expected from studies on UDD \cite{cywin08,uhrig08,bierc09a,bierc09b}, 
namely that the 
suppression of decoherence is more efficient for baths with a hard cutoff. 

\begin{figure}[ht]
    \begin{center}
    \includegraphics[width=0.99\columnwidth,clip]{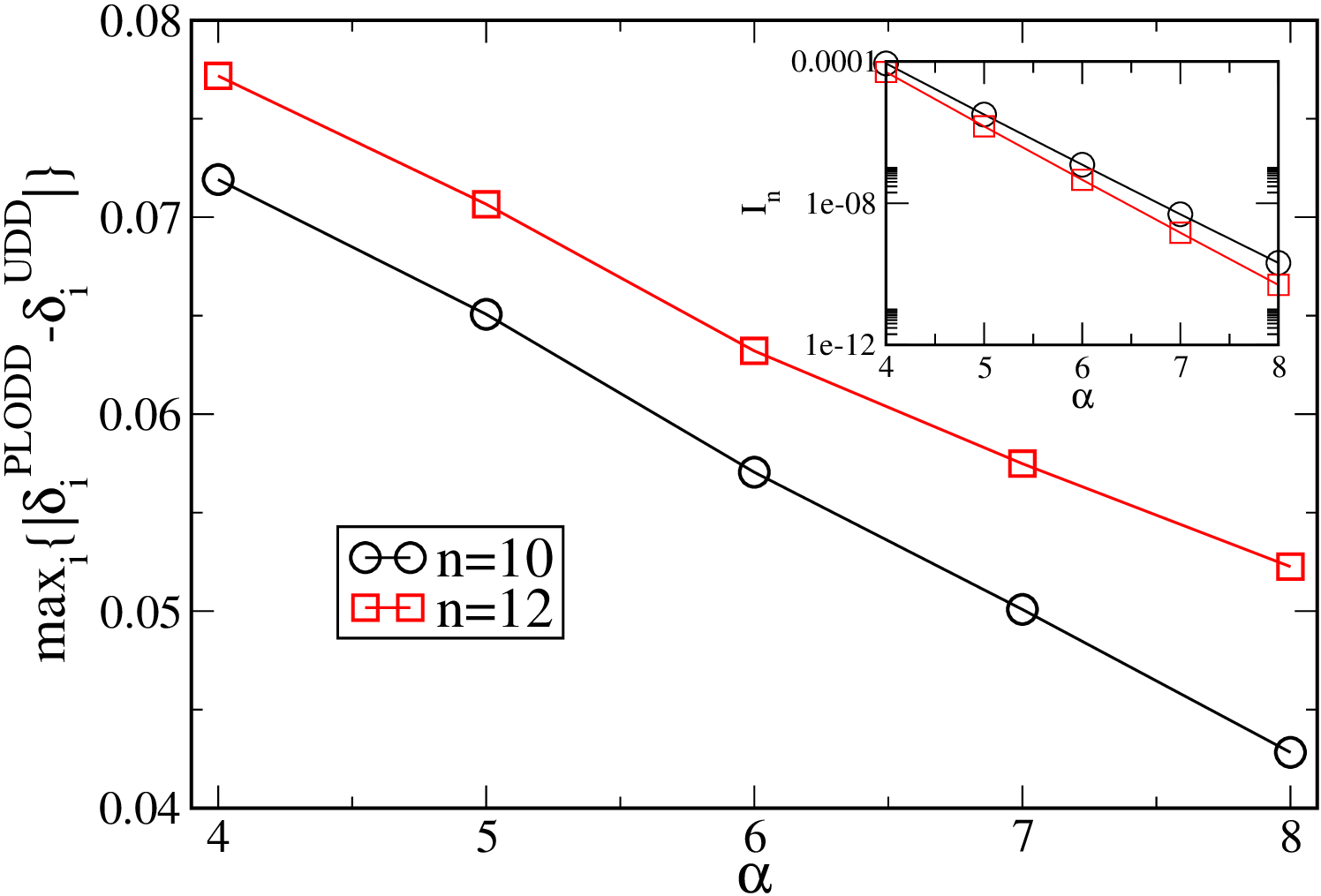}
    \end{center}
    \caption{(Color online) 
     The maximum difference between the $\{\delta_i\}$ for PLODD and UDD  
     is shown for two sequences with  $n=10$ and $n=12$ pulses.
     The PLODD sequences tend to approach the UDD sequences 
     for harder cutoffs. In
     the inset the dependence of $I_n$ on $\alpha$ 
     is depicted. Note that the $y$ axis of the inset is in logarithmic scale.
 \label{fig3}
}
\end{figure}

As $\alpha$ increases the PLODD sequences tend to coincide with the UDD
sequences. We computed  the maximum difference between the $\{\delta_i\}$ 
of the PLODD and of the UDD as a function of $\alpha$.
The results are depicted in  Fig.\ \ref{fig3}.
Clearly, the maximum difference decreases as $\alpha$ increases.
This supports that the PLODD tends to recover the UDD for large values 
of $\alpha$. This can also be seen in Fig.\ \ref{fig2_n10}. 

The inset of Fig.\ \ref{fig3}
shows that $\ln I_n$ decreases linearly with $\alpha$ for fixed values of $n$.
Combined with the power law scaling shown in Fig.\ \ref{fig_In_vs_n}
we deduce that 
\begin{equation}
\label{result2}
I_n \propto (C_{n,\alpha}/n)^{\alpha-1}
\end{equation}
holds for the PLODD sequences
where the factor $C_{n,\alpha}$ depends only very weakly on $n$
and $\alpha$.

\section{Conclusions}
\label{sec_conclusion}

We analyzed the supression of decoherence  by means of 
sequences of instantaneous $\pi$ pulses. The work horse
is the spin-boson model with pure dephasing which can be treated analytically.
The sequences are optimized for power law spectra. The 
main difference to the already known optimized sequences, for instance
UDD or OFDD,  is that the convergence of the decoherence function $\chi(t)$
is investigated for various powers ($\alpha$) of the noise spectrum.
Our approach extends previous results in several ways.

The OFDD sequences proposed by  Uys \emph{et al.\ } are optimized 
for a constant spectrum ($\alpha=1$ in our notation) with finite UV cutoff 
\cite{uys09b}. In our study we sent the UV cutoff to infinity and treated 
general power law spectra characterized by the exponent $\alpha>0$. 
Hence the proposed PLODD  sequences are optimized for arbitrary, but fixed,
exponent. They are universal in the sense that their relative
switching instants $\delta_j=t_j/t$ depend only on $\alpha$, but  not on the 
total duration $t$ of the sequence.

In their investigation of classical noise with 
exponents $\alpha\in\{1.5, 2.5\}$, Cywi\'{n}ski \emph{et al.\ }
observed that the well-known CPMG sequence works well
in the regime where the UV cutoff is infinite for
practical purposes. 
The UDD does not provide an improvement \cite{cywin08}. 
One of the authors  
generalized this investigation to the quantum mechanical
spin-boson model and cutoffs with arbitrary power law behavior \cite{uhrig08}.

In the present work we extended these
findings further by the systematic optimization of the sequence on the basis 
of analytical results for arbitrary power law spectra.
The softer the UV behavior of the power law spectrum 
is, i.e., the smaller its exponent $\alpha$ is, the more the PLODD approaches 
the CPMG.
Vice versa, the harder  the UV behavior of the power law spectrum 
is, i.e., the larger its exponent $\alpha$ is, the more the PLODD
approaches the UDD. 

Hence, the findings of previous investigations are 
corroborated. There is no completely different sequence which displays
a significantly better performance for pure dephasing other than
PLODD. The PLODD has the limiting cases CPMG (soft UV behavior) and UDD
(hard UV behavior) .

We also investigated how  the decoherence function 
$\chi(t)$ scales with $\alpha$ and $n$, the number of pulses. 
From Eqs.\ \eqref{chi_z} and \eqref{result2} we obtain 
\begin{equation}
\chi(t)\propto n(C_{n,\alpha} t/n)^\alpha,
\end{equation}
which generalizes  the result in Ref.\ \onlinecite{cywin08}
to arbitrary exponent and  a quantum mechanical model.
Here $C_{n,\alpha}$ is a factor which depends only weakly 
on $n$ and $\alpha$.

We reckon that the results of this work
represent a  useful contribution to the technique of 
dynamic decoupling. The optimization of the pulse sequence in relation to the 
specific baths is of vital importance in many applications in
high precision nuclear magnetic resonance and quantum information
processing.


\end{document}